\documentstyle[emulateapj]{article}
\submitted{Accepted for publication in the Astrophysical Journal, Letters}

\begin{document}

\title{ON THE ORIGIN OF LYMAN$\alpha$ BLOBS AT HIGH REDSHIFT:
       SUBMILLIMETRIC EVIDENCE FOR A HYPERWIND GALAXY AT $z$ = 3.1}

\author{Yoshiaki Taniguchi$^1$, Yasuhiro Shioya$^1$, \& Yuko Kakazu$^2$}

\affil{$^1$Astronomical Institute, Graduate School of Science, 
           Tohoku University, Aramaki, Aoba, Sendai 980-8578, Japan\\
       $^2$Institute for Astronomy, University of Hawaii, 2680 Woodlawn Drive,
           Honolulu, HI96822}

\begin{abstract}
The most remarkable class of high-redshift objects observed so far
is extended Ly$\alpha$ emission-line blobs found in an over-density
region at redshift 3.1. They may be either
a dust-enshrouded, extreme starburst galaxy with a large-scale galactic outflow
(superwind) or cooling radiation from dark matter halos.
Recently one of these Ly$\alpha$ blobs has
been detected at submillimeter wavelengths (450 and 850 $\mu$m).
Here we show that its rest-frame 
spectral energy distribution between optical and far-infrared
is quite similar to that of Arp 220, which is a typical ultraluminous
starburst/superwind galaxy in the local universe.
This suggests strongly that the superwind model proposed by
Taniguchi \& Shioya is applicable to this Ly$\alpha$ blob.
Since the blob is more luminous in the infrared by a factor of 30 than Arp 220,
it comprises a new population of hyperwind galaxies at high redshift.
\end{abstract}


\section{INTRODUCTION}

It was once thought that forming galaxies could be strong emission-line 
sources because a large number of massive stars cause the ionization
of gas clouds in the forming galaxies (Partridge \& Peebles 1967).
Motivated by this idea, many attempts have been made to search for such
very strong emission-line sources at high redshift, but
most these searches failed  (Pritchet 1994; Pahre \& Djorgovski 1995;
Thompson, Mannucci, \& Beckwith 1996).
However, for these past several years, Ly$\alpha$ emitters have been
found around known high-$z$ objects such as quasars
(Hu \& McMahon 1996; Hu, McMahon, \& Egami 1996; Petitjean et al. 1996; 
Hu, McMahon, \& Cowie 1999).
Further, a number of high-$z$  Ly$\alpha$ emitters have also been 
found in several blank sky areas (Pascarelle et al. 1996; Cowie \& Hu 1998;
Keel et al. 1999; Steidel et al. 2000).
These recent surveys have reinforced the potential importance of
search for high-$z$ Ly$\alpha$ emitters.

It is remarkable that two Ly$\alpha$ emitters found in 
Steidel et al. (2000) 
are observed to be very extended spatially, e.g., $\sim$ 100 kpc;
we call these Ly$\alpha$ blobs LAB1 and LAB2. 
These two LABs are found at redshift
$z \approx 3.1$ and show no evidence
for the association with AGNs\footnote{
Such extended Ly$\alpha$ blobs have been also found by
Pascarelle et al. (1996) and Keel et al. (1999).
However, since the three LABs found in K99 (53W 002, P96a Object 18,
and P96a Object 19) are all strong C {\sc iv}
emitters (Pascarelle et al. 1996), it seems natural to conclude
that they are photoionized by the central engine of active galactic
nuclei (AGNs).
Indeed, some powerful radio galaxies at high redshift
have giant Ly$\alpha$ nebulae (Chambers, Miley, \& van Bruegel 1990;
van Ojik et al. 1996).
We do not discuss these AGN-related Ly$\alpha$ blobs in this Letter.}.
Their observational properties are summarized as below
(in this Letter, we adopt an Einstein-de Sitter cosmology with
a Hubble constant $H_0 = 100 h$ km s$^{-1}$ Mpc$^{-1}$);
1) the observed Ly$\alpha$ luminosities are $\sim 10^{43} h^{-2}$
ergs s$^{-1}$, 2) they appear elongated morphologically,
3) their sizes amount to $\sim$ 100 $h^{-1}$ kpc,
4) the observed line widths amount to $\sim 1000$ km s$^{-1}$, and
5) they are not associated with strong radio-continuum sources
such as powerful radio galaxies. 

As for the origin of the LABs, two alternative ideas have been proposed.
One is that these LABs are 
superwinds driven by the initial starburst in galaxies because all the above
properties as well as the observed frequency of LABs can be explained
in terms of the superwind model (Taniguchi \& Shioya 2000).
Taniguchi \& Shioya (2000) also discussed the evolutionary link from
dust-enshrouded (or dusty) submillimeter sources (hereafter DSSs) to LABs
because the central starburst region in a forming elliptical galaxy 
could be enshrouded by a lot of gas with dust grains.
Their scenario is summarized as follows;
Step I: The initial starburst occurs in the center of pregalactic gas cloud.
Step II: This galaxy may be hidden by surrounding gas clouds for
the first several times $10^8$ years (i.e., the DSS phase).
Step III: The superwind blows and thus the DSS phase ceases.
The superwind leads to the formation of extended emission-line regions
around the galaxy (i.e., the LAB phase).
This lasts for a duration of $\sim 1 \times 10^8$ years.
And, Step IV: The galaxy evolves to an ordinary elliptical galaxy
$\sim 10^9$ years after the formation.
This superwind model predicts that the LABs are bright at 
rest-frame far-infrared if they are high-redshift, luminous analogue
of nearby superwind galaxies like Arp 220.

The other idea is that LABs are cooling radiation from proto-galaxies
or dark matter halos (Haiman, Spaans, \& Quataert 2000; Fardal et al. 2001; 
Fabian et al. 1986; Hu 1992). 
Standard cold dark matter models
predict that a large number of dark matter halos collapse at high redshift
and they can emit significant Ly$\alpha$ fluxes through collisional excitation
of hydrogen. These Ly$\alpha$ emitting halos are also consistent with 
the observed linear sizes, velocity widths, and Ly$\alpha$ fluxes of the 
LABs. However, it is uncertain how much far infrared and submillimeter
continuum emission can be emitted because little is known about the
dust content and its spatial distribution in such dark matter halos.

Very recently LAB1 has
been detected at submillimeter wavelengths (450 and 850 $\mu$m:
Chapman et al. 2001);
$S$(850 $\mu$m) = 20.1 $\pm$ 3.3 mJy and
$S$(450 $\mu$m) = 76 $\pm$ 24 mJy. 
At the position of LAB1, one Lyman break galaxy denoted 
as C11 is found in the $R$ band image. A near-infrared (NIR) source
is also found close to the center of LAB1 in the $K$ band image 
(Steidel et al. 2000)
(hereafter the source K).
If we adopt the superwind model of Taniguchi \& Shioya (2000), 
the source K is the most 
probable host galaxy giving rise to LAB1 as the relic of its superwind.

\section{THE SPECTRAL ENERGY DISTRIBUTION OF LYMAN $\alpha$ BLOB 1}

In Figure 1, we show the observed SED of LAB1. In this Figure, we show
the optical and NIR photometric data for both the source K and C11 are shown.
The submillimeter source detected at LAB1 appears indeed 
coincident with the central Ly$\alpha$ knot of LAB1 and this Ly$\alpha$
peak has the $K$-band counterpart, the source K (Chapman et al. 2001).
Therefore, the source K is the most likely host galaxy of the submillimeter
sources rather than C11.

For comparison, we show SEDs of Arp 220, NGC 6240, and SMM 02399$-$0136,
all of which are corrected as observed at $z$ = 3.1. 
Arp 220 is one of typical superwind galaxies in the local universe 
(Heckman, Armus, \& Miley 1987, 1990) 
and its infrared luminosity 
($L_{\rm IR} \equiv L_{8-1000\mu m}$) exceeds $10^{12} L_\odot$; 
i.e., a ultraluminous
infrared galaxy (ULIG: Sanders et al. 1988; Sanders \& Mirabel 1996);
$L_{\rm IR}$(Arp 220) = $1.5 \times 10^{12} L_\odot$ with a Friedman
universe with a Hubble constant of $H_0$ = 75 km s$^{-1}$ Mpc$^{-1}$.
NGC 6240 is a luminous infrared galaxy in the local universe with
evidence for an AGN (Iwasawa 1999). SMM 02399$-$0136 is a hyperluminous
infrared galaxy with evidence for an AGN at $z = 2.8$ (Ivison et al. 1998).
The observed SED of LAB1 is quite similar to that of Arp 220
although LAB1 is more luminous by a factor of 30 than Arp 220\footnote{
If we adopt a cosmology with $H_0$ = 70 km s$^{-1}$ Mpc$^{-1}$, 
$\Omega_{\rm m} = 0.3$, and $\Omega_\Lambda = 0.7$, LAB1 is more luminous
by a factor of 20 than Arp 220.}.
The SEDs of both NGC 6240 and SMM 02399$-$0136 appear different from
that of LAB1 in that they are brighter than LAB1 at rest-frame UV
and optical ranges. 
The observed $S$(450 $\mu$m)/$S$(850 $\mu$m) ratio (3.8 $\pm$ 1.2)
implies that the dust temperature is $T_{\rm dust} \simeq$ 40 K
if the source is located at $z \sim 3$ (Chapman et al. 2001; Blain 1999).
Indeed, Scoville et al. (1991) estimated 
$T_{\rm dust} \simeq$ 47 K for Arp 220, 
being consistent with the above prediction.

The comparison of SEDs shown in Figure 1 suggests strongly that
LAB1 is a very bright infrared galaxy. Since its infrared luminosity
is higher by a factor of 30 than that of Arp 220, we obtain
$L_{\rm IR}$(LAB1)  $\simeq 4.5 \times 10^{13} L_\odot$. 
Therefore, LAB1 can be regarded as a hyperluminous infrared galaxy
or a hyperwind galaxy at $z = 3.1$.

\section{DISCUSSION}

Here we consider why LAB1 is so luminous at submillimeter 
wavelengths when compared to Arp 220.
Although Arp 220 shows evidence for the superwind activity,
it is considered that ultraluminous starbursts have been going on 
for $\sim 10^8$ yr (e.g., Taniguchi, Trentham, \& Shioya 1998).
Furthermore, $\sim$ 50\% of the bolometric
luminosity is hidden by a large amount of dust grains 
(Shioya, Trentham, \& Taniguchi 2001).
This implies that the majority of the
central energy sources are still covered by a cocoon of dust grains and thus
Arp 220 is a ULIG with evidence for the superwind activity.
Therefore, it is suggested that intense star formation is still in 
progress in the heart of LAB1, although massive stars formed in earlier
evolutionary phases have already exploded as Type II supernovae,
giving rise to the formation of the observed extended Ly$\alpha$ nebula
of LAB1. 
If the origin of LAB1 is attributed to the cooling radiation from 
proto-galaxies or dark matter halos, dust grains would be associated
with distributed star formation and thus the SED of LAB1 cannot be
similar to that of Arp 220.

It is known that Arp 220 has a lot of molecular gas of $\sim 10^{10} M_\odot$
(Sanders et al. 1988; Scoville et al. 1991; Scoville, Yun, \& Bryant 1997). 
Since LAB1 appears to be a scaled-up version of Arp 220, 
it is expected that the gas mass of LAB1 may exceed $10^{11} M_\odot$,
being higher than that of a quasar host of BR 1202$-$0725 at $z = 4.7$
(Ohta et al. 1996; Omont et al. 1996). The submillimeter detection of LAB1 
implies that a large amount of dust grains are already present in this source
and thus significant flux from CO molecules will be detected in the existing
millimetric radio telescope facilities.

The superwind model for LAB1 suggests that LAB1 is a weak X-ray emitter.
Scaling the X-ray data of Arp 220 (Iwasawa 1999), we estimate a rest-frame 2 - 10 keV
flux of LAB1 (0.49 - 2.44 keV in the observed frame) as $1.22 \times 10^{-18}$ ergs
s$^{-1}$ cm$^{-2}$ for a Friedman cosmology with $H_0$ = 75 km s$^{-1}$ Mpc$^{-1}$
or  $1.74 \times 10^{-18}$ ergs
s$^{-1}$ cm$^{-2}$ for a cosmology with $H_0$ = 70 km s$^{-1}$ Mpc$^{-1}$,
$\Omega_{\rm m} = 0.3$, and $\Omega_\Lambda = 0.7$.
If this is the case, LAB1 will not be detected in the X-ray even
using {\it Chandra} (e.g., Mushotzky et al. 2000; Giacconi et al. 2001;
Tozzi et al. 2001).

Although a large number of submillimeter sources have been found 
for these past several years (Smail, Ivison, \& Blain 1997; 
Hughes et al. 1998; Barger et al. 1998; Eales et al. 1999), 
it is known that submillimeter sources at high redshift are heterogeneous 
(Ivison et al. 2000).
However, if we miss a number of hyperwind galaxies such as LAB1 in the previous
surveys, we could underestimate the star-formation rate at high redshift.
In order to elucidate the cosmic star-formation history unambiguously 
(Hughes et al. 1998; Madau et al. 1996),
it will be important to carry out deep imaging
surveys with narrow-band filters for such submillimeter sources.

\vspace{0.5cm}

We would like to thank an anonymous referee for useful suggestions
and comments.
YS is a JSPS fellow. This work was financially supported in part by
the Ministry of Education, Science, and Culture
(Nos. 10044052, and 10304013).


\begin{figure}
\epsfysize=22cm \epsfbox{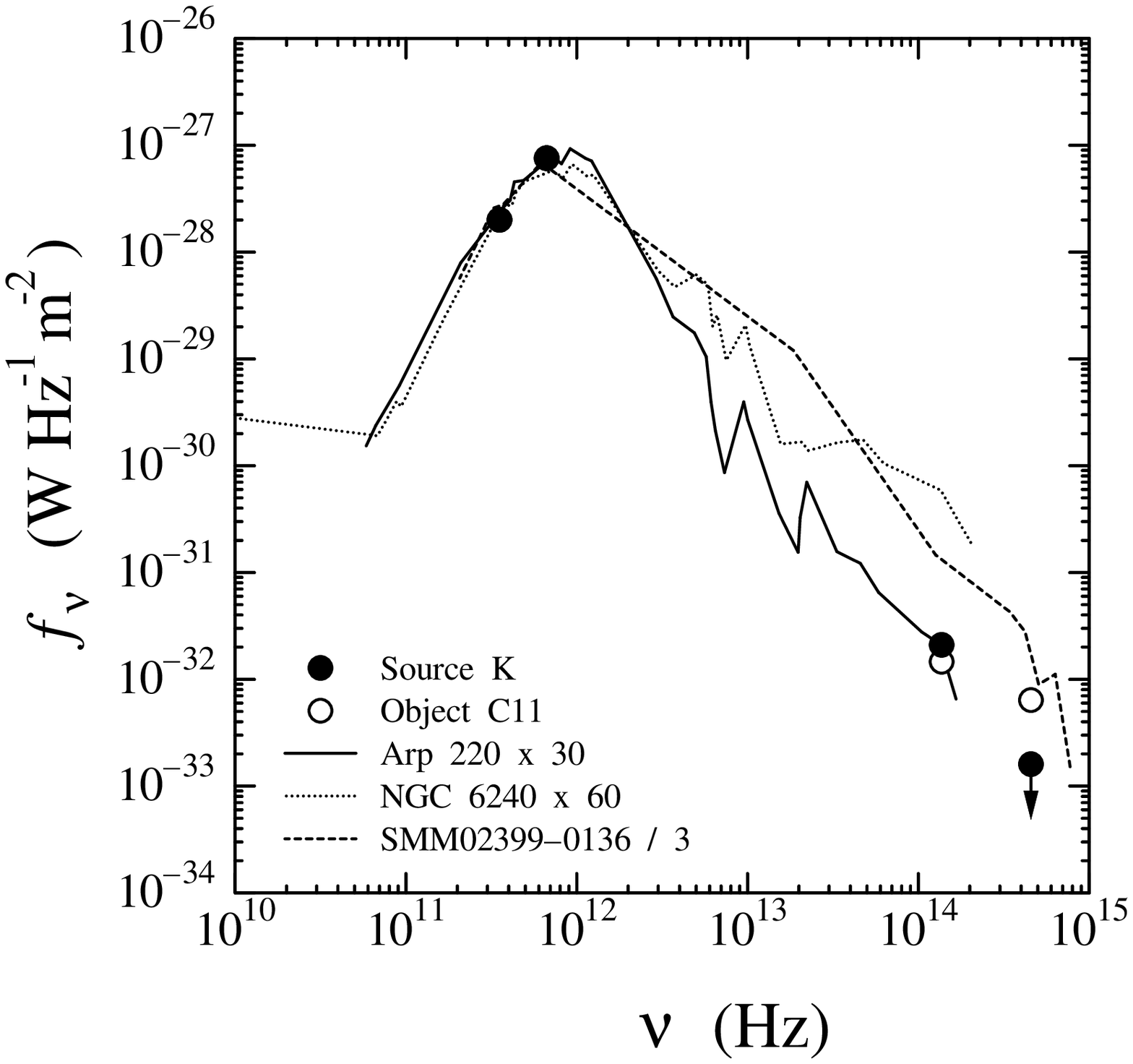}
\caption[]{
Spectral energy distribution of LAB1.
The $R$- and $K$-band photomteric data of the source K are shown by
filled circles while those of C11 are shown by open circles.
For comparison,
we show the SEDs of Arp 220 (solid line), NGC 6240 (dotted line),
and SMM 02399$-$0136 (dashed line). Their photometric data are
taken from Rigopoulou, Lawrence \& Rowan-Robinson (1996) and
Klaas et al. (1997) for Arp 220, Rigopoulou et al. (1996),
Klaas et al. (1997), Griffith et al. (1995), Lisenfeld, Isaak \& Hills
(2000), Spinoglio et al. (1995) and White \& Becker (1992) for NGC 6240,
and Ivison et al. (1998) for SMM 02399$-$0136.
\label{fig1}
}
\end{figure}


\begin{references}
\reference{1}{Barger, A.,  Cowie, L. L., Sanders, D. B., Fulton, E.,
              Taniguchi, Y., Sato, Y., Kawara, K., \& Okuda, H.
              1998, Nature, 394, 248}
\reference{1}{Blain, A. 1999, MNRAS, 309, 955}
\reference{1}{Chambers, K. C., Miley, H. K. \& van Bruegel, W. J. M. 1990,
                  ApJ, 363, 21}
\reference{1}{Chapman, S. C., Lewis, G. F., Scott, D., Richards, E.,
              Borys, C., Steidel, C. C., Adelberger, K. L., \& Shapley, A. E.
              2001, ApJ, 548, L17}
\reference{1}{Cowie, L. L., \& Hu, E. M. 1998, AJ, 115, 1319}
\reference{1}{Eales, S., Lilly, S., Gear, W., Dunne, L., Bond, J. R., Hammer, F.,
              Le F\`evre, O., \& Crampton, D. 1999, ApJ, 515, 518}
\reference{1}{Fabian, A. C., Arnaud, K. A., Nulsen, P. E. J., \&
              Mushotzky, R. F. 1986, ApJ, 305, 9}
\reference{1}{Fardal, M. A., Katz, N., Gardner, J. P., Hernquist, L., Weinberg, D. H.,
              \& Dav\'e, R. 2001, ApJ, in press (astro-ph/0007205)}
\reference{1}{Giacconi, R., et al. 2001, ApJ, 551, 624}
\reference{1}{Griffith, M. R., Wright, A. E., Burke, B. F., \& Ekers, R. D.
              1995, ApJS, 97, 347}
\reference{1}{Haiman, Z., Spaans, M., \& Quataert, E. 2000, ApJ, 537, L5}
\reference{1}{Heckman, T. M., Armus, L., \& Miley, G. K. 1987, AJ, 93, 276}
\reference{1}{Heckman, T. M., Armus, L., \& Miley, G. K. 1990, ApJS,
              74, 833}
\reference{1}{Hu, E. M. 1992, ApJ, 391, 608}
\reference{1}{Hu, E. M., \& McMahon, R. G. 1996, Nature, 382, 231}
\reference{1}{Hu, E. M., McMahon, R. G., \& Cowie, L. L. 1999, ApJ, 522, L9}
\reference{1}{Hu, E. M., McMahon, R. G., \& Egami, E. 1996, ApJ, 459, L53}
\reference{1}{Hughes, D., et al. 1998, Nature, 394, 241}
\reference{1}{Ivison, R. J., Smail, I., Barger, A. J., Kneib, J. -P., Blain,
              A. W., Owen, F. N., Kerr. T. H., \& Cowie, L. L. 2000, MNRAS, 315, 209}
\reference{1}{Ivison, R. J., Smail, I., Le Borgne, J. -F., Blain, A. W.
              Kneib, J. -P., B\'ezencourt, J., Kerr, T. H., \& Davies, J. K.
              1998, MNRAS, 298, 583}
\reference{1}{Iwasawa, K. 1999, MNRAS, 302, 96}
\reference{1}{Keel, W. C., Cohen, S. H., Windhorst, R. A., \& Waddington,
              I. 1999, AJ, 118, 2547 (K99)}
\reference{1}{Klaas, U., Haas, M., Heinrichsen, I., \& Schultz, B. 1997,
              A\&A, 325, L21}
\reference{1}{Lisenfeld, U., Isaak, K. G., \& Hills, R. 2000, MNRAS, 312, 433}
\reference{1}{Madau, P., Ferguson, H. C., Dickinson, M. E., Giavalisco, M.,
              Steidel, C. C., \& Fruchter, A. 1996, MNRAS, 283, 1388}
\reference{1}{Mushotzky, R. F., Cowie, L. L., Barger, A. J., \& Arnaud, L. A.
              2000, Nature, 404, 459}
\reference{1}{Ohta, K. Yamada, T., Nakanishi, K., Kohno, K., Akiyama, M.,
                  \& Kawabe, R. 1996, Nature, 382, 426}
\reference{1}{Omont, A., Petitjean, P., Guilloteau, S., McMahon, R. G., 
                  Solomon, P. M., \& P\'econtal, E. 1996, Nature, 382, 428}
\reference{1}{Pahre, M. A., \& Djorgovski, S. D. 1995, ApJ, 449, L1}
\reference{1}{Partridge, R. B., \& Peebles, P. J. E. 1967, ApJ, 147, 868}
\reference{1}{Pascarelle, S. M., Windhorst, R. A., Keel, W. C., \&
              Odewahn, S. C. 1996, Nature, 383, 45}
\reference{1}{Petitjean, P., P\'econtal, E., Valls-Gabaud, D., \&
              Charlot, S. 1996, Nature, 380, 411}
\reference{1}{Pritchet, C. J. 1994, PASP, 106, 1052}
\reference{1}{Rigopoulou, D., Lawrence, A., \& Rowan-Robinson, M. 1996,
              MNRAS, 278, 1049}
\reference{1}{Sanders, D. B., \& Mirabel, I. F. 1996, ARA\&A, 34, 749}
\reference{1}{Sanders, D. B., et al. 1988, ApJ, 325, 74}
\reference{1}{Scoville, N. Z., Yun, M. S., \& Bryant, P. M. 1997, ApJ, 484, 702}
\reference{1}{Scoville, N. Z., Sargent, A. I., Sanders, D. B., \& Soifer, 
              B. T. 1991, ApJ, 366, L5}
\reference{1}{Shioya, Y., Trentham, N., \& Taniguchi, Y. 2001, ApJ, 548, L29}
\reference{1}{Smail, I., Ivison, R. J., \& Blain, A. W. 1997, ApJ, 490, L5}
\reference{1}{Spinoglio, L., Malkan, M. A., Rush, B., Carrasco, L, \& Recillas-Cruz, E.
              1995, ApJ, 453, 616}
\reference{1}{Steidel, C. S., Adelberger, K. L., Shapley, A. E., Pettini,
              M., Dickinson, M., \& Giavalisco, M. 2000, ApJ, 532, 170}
\reference{1}{Taniguchi, Y., \& Shioya, Y. 2000, ApJ, 532, L13}
\reference{1}{Taniguchi, Y.,  Trentham, N., \& Shioya, Y. 1998, ApJ, 504, L79}
\reference{1}{Thompson, D., Mannucci, F., \& Beckwith, S. V. W. 1996,
                 AJ, 112, 1794}
\reference{1}{Tozzi, P., et al. 2001, ApJ, in press (astro-ph/0103014)}
\reference{1}{van Ojik, R., R\"ottgering, H. J. A., Carilli, C. L., Miley, G. K.,
                  \& Bremer, M. N. 1996, A\&A, 313, 25}
\reference{1}{White, R. L., \& Becker, R. H. 1992, ApJS, 79, 331}
\end{references}
\end{document}